\documentclass[12pt,preprint]{aastex}
 
\usepackage{graphicx}
\usepackage{psfig}

\def\deg      {{\ifmmode^\circ\else$^\circ$\fi}} 

\slugcomment{Resubmitted to COSMOS ApJ Special Issue October 16, 2006}
 
 \shorttitle{Radio and millimeter properties of $z \sim 5.7$ 
Ly$\alpha$ emitters in the COSMOS field}
\shortauthors{Carilli et al.}
 
\begin{document}

 \title{Radio and millimeter properties of $z \sim 5.7$ 
Ly$\alpha$ emitters in the COSMOS field:
limits on radio AGN, submm galaxies, and dust obscuration}
 
 \author{ 
C.L. Carilli\altaffilmark{1},
T. Murayama\altaffilmark{2},
R. Wang\altaffilmark{1},
E. Schinnerer\altaffilmark{3},
Y. Taniguchi\altaffilmark{2},
V. Smol{\v c}i{\' c}\altaffilmark{3},
F. Bertoldi\altaffilmark{4},
M. Ajiki\altaffilmark{2},
T. Nagao\altaffilmark{5},
S. S. Sasaki\altaffilmark{6},
Y. Shioya\altaffilmark{2},
J.~E.~Aguirre\altaffilmark{7,8}
A. W. Blain\altaffilmark{9},
N. Scoville\altaffilmark{9},
D. B. Sanders\altaffilmark{10}
}

\altaffiltext{$\star$}{Based on data collected at : the Subaru
Telescope, which is operated by the National Astronomical Observatory
of Japan; the National Radio Astronomy Observatory which is a facility
of the National Science Foundation operated under cooperative
agreement by Associated Universities, Inc; the IRAM 30m telescope; the
Caltech Submm Observatory}

\altaffiltext{1}{National Radio Astronomy Observatory, P.O. Box 0, Socorro, NM
87801}
\altaffiltext{2}{Physics Department, Graduate School of Science \& 
    Engineering, Ehime University, 2-5 Bunkyo-cho, Matsuyama
    790-8577, Japan}
\altaffiltext{3}{Max Planck Institut f\"ur Astronomie, K\"onigstuhl
  17, Heidelberg, D-69117, Germany}
\altaffiltext{4}{Dept. Astronomy, Bonn University, Bonn, Germany, D53121}
\altaffiltext{5}{National Astronomical Observatory of Japan,
          2-21-1 Osawa, Mitaka, Tokyo 181-8588, Japan}
\altaffiltext{6}{Astronomical Institute, Graduate School of Science,
         Tohoku University, Aramaki, Aoba, Sendai 980-8578, Japan}
\altaffiltext{7}{Jansky Fellow, National Radio Astronomy Observatory}
\altaffiltext{8}{Center for Astrophysics and Space Astronomy and
Department of Astronomy and Planetary Sciences, University of
Colorado, 593 UCB, Boulder, CO 80303-0593}
\altaffiltext{9}{California Institute of Technology, MC 105-24, 1200 East
California Boulevard, Pasadena, CA 91125}
\altaffiltext{10}{Institute for Astronomy, 2680 Woodlawn Dr.,
  University of Hawaii, Honolulu, Hawaii, 96822}

\begin{abstract}

We present observations at 1.4 and 250 GHz of the $z\sim 5.7$
Ly$\alpha$ emitters (LAE) in the COSMOS field found by Murayama et
al.. At 1.4 GHz there are 99 LAEs in the lower noise regions of the
radio field.  We do not detect any individual source down to 3$\sigma$
limits of $\sim 30\mu$Jy beam$^{-1}$ at 1.4 GHz, nor do we detect a
source in a stacking analysis, to a 2$\sigma$ limit of $2.5\mu$Jy
beam$^{-1}$.  At 250 GHz we do not detect any of the 10 LAEs that are
located within the central regions of the COSMOS field covered by
MAMBO ($20' \times 20'$) to a typical 2$\sigma$ limit of $S_{250} <
2$mJy.  The radio data imply that there are no low luminosity radio
AGN with $L_{1.4} > 6\times 10^{24}$ W Hz$^{-1}$ in the LAE
sample. The radio and millimeter observations also rule out any highly
obscured, extreme starbursts in the sample, ie. any galaxies with
massive star formation rates $> 1500$ M$_\odot$ year$^{-1}$ in the
full sample (based on the radio data), or 500 M$_\odot$ year$^{-1}$
for the 10\% of the LAE sample that fall in the central MAMBO
field. The stacking analysis implies an upper limit to the mean
massive star formation rate of $\sim 100$ M$_\odot$ year$^{-1}$.

\end{abstract}
 
 \keywords{galaxies: formation --- galaxies: evolution --- 
galaxies: radio, submm, IR --- surveys}

\section{Introduction}

Numerous studies have demonstrated the power of discovering high
redshift star forming galaxies using narrow band filters centered on
the Ly$\alpha$ line (Hu et al. 2002, 2004, Kodaira et al. 2003, Rhoads
et al. 2003, Malhotra \& Rhoads 2004, Tran et al. 2004, Kurk et
al. 2004, Santos et al. 2004, Martin \& Sawicki 2004, Taniguchi et
al. 2005; Iye et al. 2006).  Indeed, the majority of galaxies known at
$z \sim 6$ have been discovered in this way.  Finding galaxies at
these extreme redshifts is of paramount importance since the recent
discovery of Gunn-Peterson absorption by a partially neutral IGM
toward the highest $z$ QSOs ($z \sim 6$; Fan et al. 2006) -- a
signature of cosmic reionization.  Reionization is a key benchmark in
cosmic structure formation, indicating the formation of the first
luminous objects (Fan, Carilli, Keating 2006).

The Cosmic Evolution Survey (COSMOS), covering 2 \sq\deg, is designed
to probe the evolution of galaxies, AGN and dark matter in the context
of their cosmic environment. The COSMOS/HST field has extensive
supporting observations, ranging from the radio through the X-ray
(Scoville et al. 2006).  Part of this program entails a SUBARU narrow
band survey of the full field centered on Ly$\alpha$ at $z \sim 5.7$
(Murayama et al. 2006). This survey has revealed a large sample of
galaxies at $z \sim 5.7$, with 110 candidate galaxies.

Observations of the COSMOS field have been done at 1.5$"$ resolution
(FWHM) at 1.4 GHz down to an rms level between 8 and 10$\mu$Jy
beam$^{-1}$ (Schinnerer et al.  2006).  Observations have also been
done at 250 GHz at a resolution of 10.6$''$ of the inner $20'\times
20'$ of the COSMOS field using MAMBO at the IRAM 30m telescope to a
rms level of 0.9 mJy (Bertoldi et al 2006), and a somewhat larger,
shallower field ($30.6'\times 30.6'$) at a resolution of 31$''$ using
BOLOCAM at the Caltech Submm Observatory (Aguirre et al. 2006) to an
rms level of 1.9mJy.

In this paper we use the data from the VLA, MAMBO, and BOLOCAM of the
COSMOS field to constrain the centimeter and millimeter properties of
the $z=5.7$ LAEs.  This study represents the deepest radio continuum
study of high-$z$ LAEs, over the largest area, as well as the most
extensive study of these sources at millimeter wavelengths to date.
These data allow us to set limits on any low luminosity radio AGN, as
well as on the number of highly dust-obscured starburst galaxies, in
the LAE sample at $z=5.7$.  We perform a stacking analysis to set a
limit to the mean UV obscuration of high-$z$ LAEs.
 
\section{The sample and the radio and millimeter observations}

\subsection{The sample} 

The sample is taken from the narrow band Ly$\alpha$ survey of Murayama
et al. (2006) centered on a redshift of $z=5.7 \pm 0.05$.  They cover
the full COSMOS field, implying a comoving volume of $1.5\times10^6$
Mpc$^{3}$.  They select sources that are detected in the narrow band
NB816 filter at NB816 $< 25.1$ mag, are undetected in shorter
wavelength broad band filters, and have NB-to-broad band near-IR
colors that imply Ly-$\alpha$ (observed) equivalent widths, $EW_{obs}
> 120 \AA$ (corresponding to rest frame $EW_{rest} = EW_{obs}/(1+z) >
18\AA$).

They find 110 candidate LAEs, and they estimate that the
contamination rate by low $z$ objects is $< 14\%$.  Thirty seven
sources are also detected in longer wavelength filters ($z'$),
corresponding to rest frame UV emission (1250\AA).
All of the sources are small, $< 0.5"$, and a few ($\sim 5\%$) show
evidence for 2 or 3 compact components. No large ($\ge 10$'s kpc) 
'Ly-$\alpha$ blob'  sources are detected (Steidel et al. 2000;
Matsuda et al. 2004).

\subsection{The VLA observations}

We have searched for radio emission from the LAEs in the COSMOS
field using the data presented in Schinnerer et al. (2006).  At each
position we determine the flux density, and the rms noise in the
region.  The relative astrometric accuracy between the radio and
optical images is better than 0.2$"$ (Aussel et al. 2006), while for a
3$\sigma$ detection the positional uncertainty is given roughly by:
FWHM/SNR $\sim 0.5''$. We have searched for radio sources
within 0.6$''$ radius of the LAE optical position.
We exclude from the analysis 11 LAEs in higher noise regions
of the field, such as close to a bright continuum source, or
near the edge of the field, leaving a sample of 99 sources total,
and 33 with UV continuum detections.

We do not detect any source $>3\sigma$ at 1.4 GHz within 0.6$''$ of
any LAE in the sample of 99. The typical 3$\sigma$ limit is 30$\mu$Jy
beam$^{-1}$ at 1.4GHz.  Note that, for the full sample of 99 sources,
we expect 0.05 chance coincidences within 0.6$''$ at the level of
30$\mu$Jy beam$^{-1}$, based on faint radio source counts (Fomalont et
al. 2006, in prep). 
 
We have also performed a radio stacking analysis of the sources,
summing images centered on the positions of the LAEs, weighted
by the rms in each subfield.  Stacking all the LAEs, we do
not detect a source at the LAE position to a $2\sigma$ limit of
2.5$\mu$Jy beam$^{-1}$. If we only stack the UV-detected sources (33
sources), we find a $2\sigma$ limit of $4 \mu$Jy beam$^{-1}$ (Figure
1).

One LAE, J10000.51+014940.1, has a marginal ($2.7\sigma$) 1.4 GHz
source of $27\pm 10\mu$Jy located just $0.2''$ from the optical
position (Figure 2). If real, the implied luminosity density at a rest
frame frequency of 1.4 GHz is $L_{1.4} = 6\times 10^{24}$ W
Hz$^{-1}$, assuming a spectral index of --0.75.  We do not consider
this a firm detection, but deeper radio imaging would be very
interesting for this source.

We note that in the study of the GOODS North field, Ajiki et al. (2006) 
found 10 LAEs at $z \sim 5.7$ using a similar technique as was employed
for the COSMOS field. Comparing these sources to the deep radio survey
of Richards (2000), we again find that no LAE has a radio counterpart 
to a 5$\sigma$ detection limit of $40\mu$Jy.

\subsection{The MAMBO and BOLOCAM observations}

Ten of the LAEs are located within the $20'\times 20'$ field imaged
with MAMBO at 250 GHz (Bertoldi et al. 2006).  Given a FWHM of
10.6$''$, we searched for MAMBO counterparts $> 3\sigma$, within
3.5$''$ of an LAE.  None of the LAEs have a MAMBO counterpart, to a
typical 3$\sigma$ upper limit of $S_{250} < 3$ mJy.  A stacking
analysis leads to a $2\sigma$ limit to the mean 250 GHz flux density
of $S_{250} < 0.7$mJy.  For reference, based on submm galaxy source
counts (Bertoldi et al. 2006), we expect 0.01 chance coincidences
within 3.5$''$ with $S_{250} \ge 3$mJy for the 10 LAEs located in the
MAMBO-COSMOS field.

There is one LAE, J100040.22+021903.8, that has a potential MAMBO
source located 5$"$ south of the LAE position, with a flux density of
$S_{250} = 3.2 \pm 0.91$ mJy (Figure 3). The BOLOCAM image shows a
value of $1.7\pm 1.9$ mJy at this position. The radio image shows a
surface brightness of $15\pm10\mu$Jy beam$^{-1}$ at the LAE position.
This LAE is not detected in the UV continuum, and the total star
formation rate based on the Ly$\alpha$ luminosity is only 6 M$_\odot$
year$^{-1}$ (uncorrected for obscuration).  Given the positional
offset, and the relatively low significance of the MAMBO detection, we
cannot claim either reality of the MAMBO source, or an association of
the LAE and the (marginal) MAMBO source.  If real, the implied FIR
luminosity is: $L_{FIR} = 1.1\times10^{13}$ L$_\odot$, and the
predicted radio flux density at 1.4 GHz is 10$\mu$Jy based on a star
forming galaxy template (Carilli \& Yun 2000). Deeper observations at
250 and 1.4 GHz are required to check the reality of this source.

We also searched the wider, shallower BOLOCAM field for counterparts
to the LAEs.  There are 12 LAEs in the BOLOCAM field (10 are common to 
the MAMBO field), and again, no source is detected with BOLOCAM to
a typical 3$\sigma$ limit of 5.7mJy. A stacking analysis provides a
2$\sigma$ limit of 1 mJy. 

\section{Discussion}

We do not detect any individual source to a typical 3$\sigma$ limit of
30$\mu$Jy beam$^{-1}$ at 1.4 GHz.  A limit of 30$\mu$Jy beam$^{-1}$ at
an observing frequency of 1.4 GHz implies a limit to the radio
luminosity at an emitted frequency of 1.4 GHz of $L_{1.4} <
6\times10^{24}$ W Hz$^{-1}$, assuming a spectral index of $-0.75$.
For comparison, the nearby Fanaroff-Riley class I (ie. low luminosity)
radio galaxy M87 has $L_{1.4} = 9\times10^{24}$ W Hz$^{-1}$.  The lack
of radio AGN in the LAE sample is not surprising, since Taniguchi et
al. (2006, in prep) show that the narrow band search technique selects
against broad line QSOs, for which the emission lines are typically
broader than the filter. Likewise, Hu et al. (1998) and Keel et
al. (1999) find a relatively low fraction (between 17\% and 40\%) of
narrow line AGN in lower $z$ LAE samples, while Shapley et al. (2003)
find only 3\% of the Ly-break galaxies at $z \sim 3$ show optical
emission line spectra consistent with an AGN.  Overall, our
non-detection of even a low luminosity radio AGN in any of the 99
COSMOS LAEs is broadly consistent with the conclusion that the narrow
band Ly$\alpha$ search technique preferentially selects for
star-forming galaxies.

We should point out that, in their extensive study of a sample of NB
selected LAEs at $z \sim 4.5$, Malhotra \& Rhoads (2002) found a
surprising fraction of the sources ($\sim 60\%$) had Ly-$\alpha$
$EW_{rest} > 240\AA$. They state that such large EW's cannot arise
through normal star formation, requiring either: (i) a narrow-line
AGN, (ii) a top-heavy IMF, or (iii) low metalicities.  The largest
measured $EW_{rest}$ in the COSMOS LAE sample is 103\AA, however, the
majority of sources are not detected in the UV, and hence only lower
limits on the $EW$ values can be set (see Section 2.1). Murayama et
al. (2006) discuss the $EW$ distribution for the COSMOS sample in more
detail.

The radio luminosity limit also corresponds to a massive ($>
5$M$_\odot$) star formation rate $\sim 1500$ M$_\odot$ year$^{-1}$
(Condon 1992).  Hence, we can rule-out any highly dust obscured,
'hyperluminous' infrared starburst galaxy.  Such a source would
correspond to a bright 'submm' galaxy with a 250 GHz flux density of
$\sim 9$mJy, assuming that the local far-IR--radio correlation
continues to apply to redshift $z\simeq 6$ (eg.  Blain et al. 2001;
Bertoldi et al. 2006; Carilli \& Yun 1999, 2000).  The MAMBO image of
the inner 20$'$ of the COSMOS field pushes this limit down to 3mJy
(500 M$_\odot$ year$^{-1}$), at least for the 10\% of the LAE sample
that fall within this area.

The mean total (0.1 to 100 M$_\odot$) star formation rate for all the
sources based on the Ly-$\alpha$ luminosity is $\sim 8$ M$_\odot$
year$^{-1}$ (Murayama et al. 2006). A similar number is found for the
star formation rates derived from the Ly-$\alpha$ luminosity for the
UV-detected subsample. For comparison, the star formation rates
derived from UV luminosities are systematically higher, with the mean
total star formation rate derived from the UV luminosities for the
UV-detected sources $\sim 12$ M$_\odot$ year$^{-1}$.  The implied
massive star formation rates (5 to 100 M$_\odot$), assuming a
Salpeter IMF, are a factor 5.6 smaller, or 1.4 and 2.1 M$_\odot$
year$^{-1}$, respectively.  The difference between the Ly-$\alpha$
derived and UV luminosity derived star formation rates is discussed
in Murayama et al. (2006), and likely relates to extra attenuation of
the Ly-$\alpha$ line due to associated Ly-$\alpha$ absorption.  Note
that none of these values have been corrected for dust extinction.

From the radio stacking analysis, we derive a (2$\sigma$) upper limit
to the mean massive star formation rate of 81 M$_\odot$ year$^{-1}$
for all the LAEs, and 130 M$_\odot$ year$^{-1}$ for just the
UV-detected sources.  These radio limits to the star formation rate
are independent of the dust content.  Hence, the upper limit to the
mean obscuration of the LAE galaxies in either the UV continuum or the
Ly-$\alpha$ line, is about factor of 60.  For comparison, the typical
Ly-break galaxy is thought to have its UV emission attenuated by a
factor $\sim 5$ due to intrinsic dust (Steidel et al. 1999), and the
mean obscuration for galaxies selected using the Ly-$\alpha$ narrow
band technique is thought to be even smaller (Shapely et
al. 2003). Hence, while our study represents the most sensitive,
widest field radio and mm study of high-$z$ LAEs to date, it also
accentuates the relatively poor limits that can be reached in the
radio and mm for star forming galaxies at the highest redshifts, when
compared to studies using the Ly$\alpha$ line.

The main result of this work is to rule-out the existence of any
highly obscured massive starburst, or low luminosity radio AGN in the
COSMOS LAE sample. Clearly, to push down to normal star forming
galaxies will require the one to two orders of magnitude improvement
in sensitivity afforded by the up-coming Expanded Very Large Array,
and the Atacama Large mm Array (ALMA).

\acknowledgments

CC acknowledges partial support from the Max-Planck Society and the
Alexander von Humboldt Foundation through the Max-Planck
Forshungspreise 2005.  We thank the referee for very useful comments.
The HST COSMOS Treasury program was supported through NASA grant
HST-GO-09822. We gratefully acknowledge the contributions of the
entire COSMOS collaboration.  More information on the COSMOS survey is
available at {\bf \url{http://www.astro.caltech.edu/cosmos}}.

 \clearpage\newpage

\noindent{\bf Note Added in Proof}

Subsequent to the acceptance of the paper: "Radio and submm
observations of LAEs in the Cosmos field," by Carilli et al., an
additional nine LAEs were discovered on further investigation of the
Subaru images, making for a total of 119 LAEs in the Cosmos field
(Murayama et al. 2006). We have searched the radio and submm Cosmos
images for counterparts, and do not detect any source to similar
limits as those presented for the original sample of 110 LAEs. The
stacking analysis is effectively unaltered by these new sources.

There is one source in the new sample of nine (J095825.26+022651.32 =
source 4 in the final LAE catalog; Murayama et al. 2006) which
projects within 5$"$ north-west of a strong radio hot spot.  This
radio hot spot is at the end of one of the radio lobes of an
arcminute-sized luminous radio galaxy. The (likely) optical
identification of the radio host galaxy is an early-type galaxy with a
photometric redshift of $1.1\pm 0.2$, situated in a cluster of
galaxies at this redshift identified by Finoguenov et al. (2006).  We
feel the projected proximity of the LAE and the radio hot spot is most
likely just a coincidence, although it is possible that gravitational
lensing by the cluster may magnify the LAE, or that the detected
excess in the NB816 filter is due to broad [OII] 372.7nm nebular
emission at $z=1.19$, associated with shocked gas preceding the radio
hot spot. Spectroscopy of this object is needed to test these
possibilities.

\clearpage
\newpage
 
\begin{figure}[ht]
\psfig{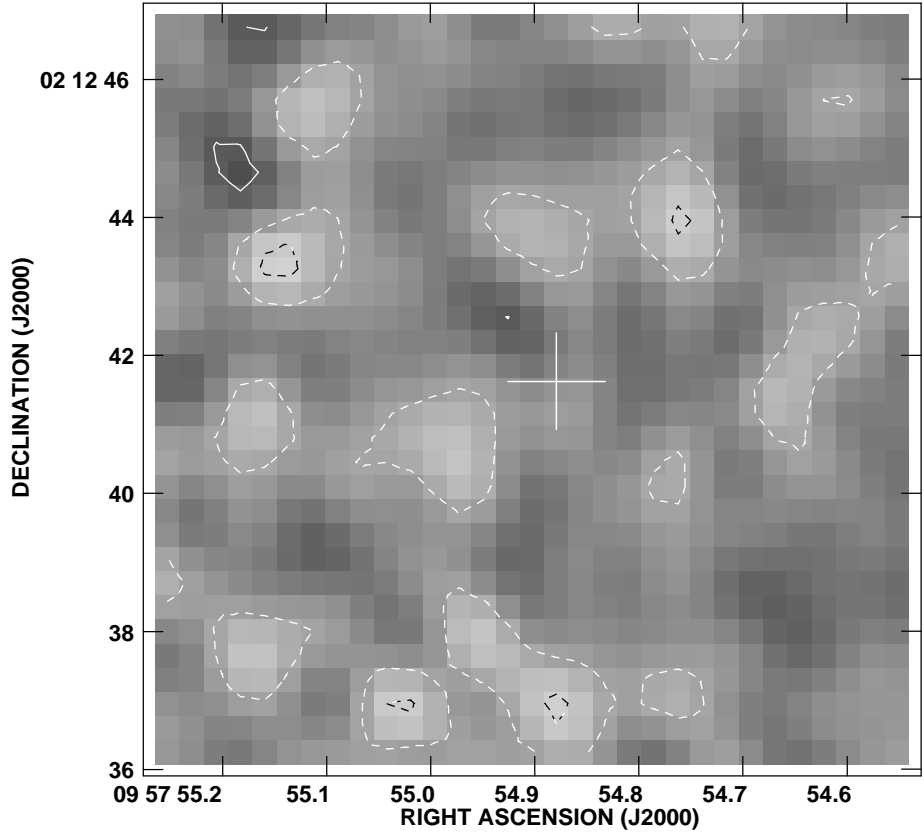}
\caption{The stacked 1.4 GHz image of the 
LAEs in the COSMOS field (99 sources). The rms noise level is 1.25 $\mu$Jy
beam$^{-1}$. The cross marks the stacking position, centered
on the LAE positions (the absolute coordinates axes are arbitrary).   
The contour levels are: -6, -4, -2, 2, 4, 6 $\mu$Jy beam$^{-1}$, 
and the beam has FWHM = 1.5$"$.
} 
\label{}
\end{figure}

 \clearpage\newpage

\begin{figure}[ht]
\psfig{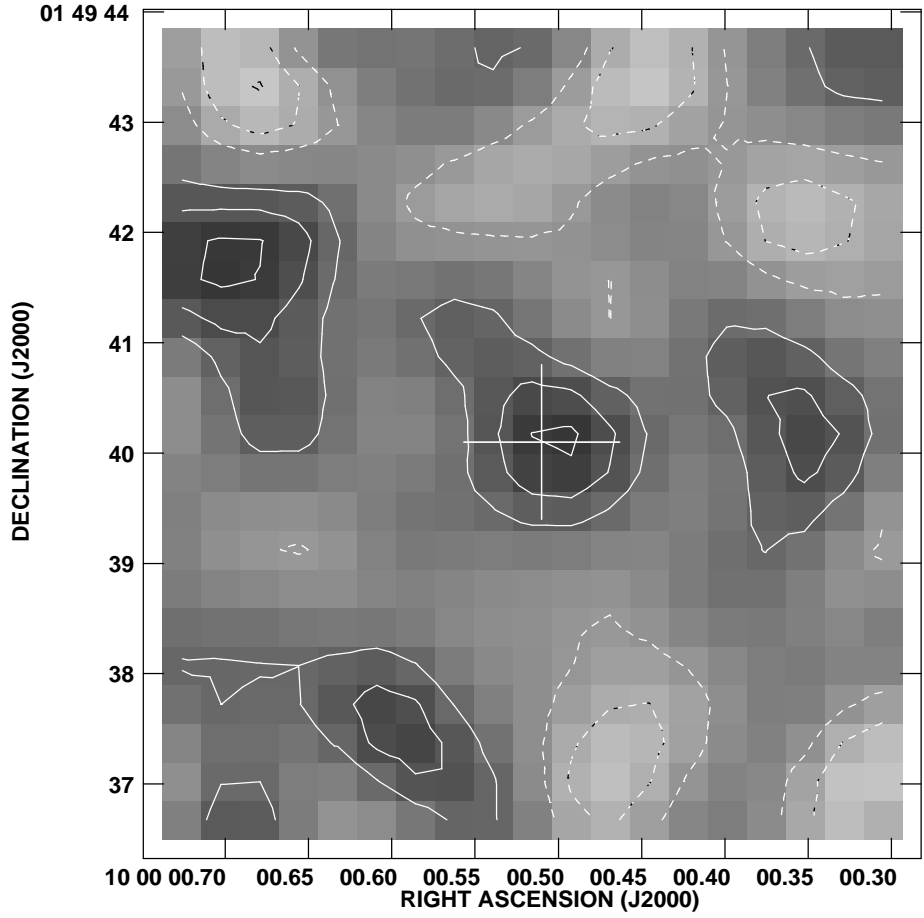}
\caption{The VLA 1.4 GHZ image of the field centered on 
the $z=5.7$ LAE J10000.51+014940.1  in Ajiki et al. (2006). 
The cross marks the position of the LAE. 
The contour levels are: -27, -18, -9, 9, 18, 
27 $\mu$Jy beam$^{-1}$,  and the beam has FWHM = 1.5$"$.
} 
\label{}
\end{figure}

 \clearpage\newpage
 
\begin{figure}[ht]
\psfig{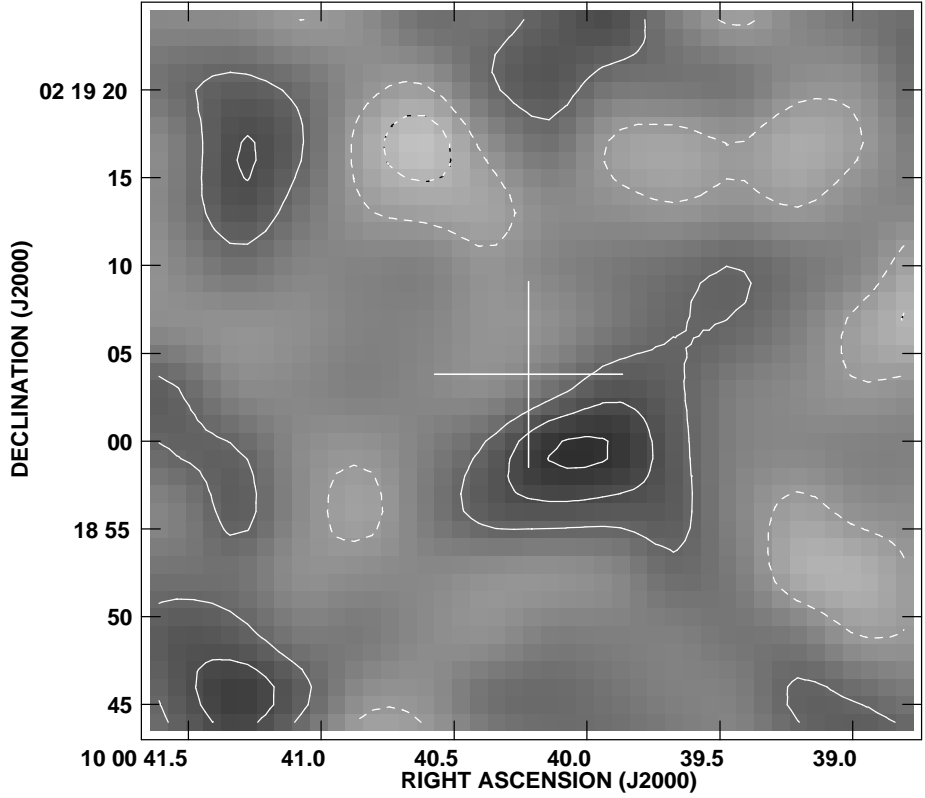}
\caption{The MAMBO 250 GHz image of the field centered on 
the $z=5.7$ LAE J100040.22+021903.8 in Ajiki et al. (2006). 
The cross marks the position of the LAE, and the 
size of the cross corresponds to the FWHM of the 
MAMBO beam. The contour levels are: -3, -2, -1, 1, 2, 3 mJy
beam$^{-1}$,  and the beam has FWHM = 10.6$''$.
} 
\label{}
\end{figure}
 
\end{document}